\documentclass[12pt]{article}
\usepackage{amssymb}
\usepackage{amsmath}
\usepackage{hyperref}
\begin{document}
\title{\bf The effect of three matters on KSS bound }
\author{ %Shahrokh Parvizi$^1$\thanks{Corresponding author: Email:parvizi@modares.ac.ir}  \hspace{2mm} and
	Mehdi Sadeghi\thanks{Email:  mehdi.sadeghi@abru.ac.ir}\hspace{2mm}\\
	%	{\small {\em $^1$Department of Physics, School of Sciences,}}\\
	%   {\small {\em Tarbiat Modares University, P.O.Box 14155-4838, Tehran, Iran }}\\
	{\small {\em Department of Physics, School of Sciences,}}\\
	{\small {\em Ayatollah Boroujerdi University, Boroujerd, Iran}}
}
\date{\today}
\maketitle

\abstract{In this paper we introduce the black brane solutions in AdS space in 4-dimensional (4D) Einstein-Gauss-Bonnet-Yang-Mills theory in the presence of string cloud and quintessence. Shear viscosity to entropy density ratio is computed via fluid-gravity duality, as a transport coefficient for this model.}\\

%\pacs{11.10.Jj, 11.10.Wx, 11.15.Pg, 11.25.Tq}


%--------------------------------------------------------------------------
\section{Introduction} \label{intro}

\indent The AdS/CFT duality \cite{Maldacena}-\cite{Witten:1998qj} originates from string theory and provides a new perspective on quantum gravity. In its most general form, it is known as a gauge/gravity duality. This duality relates weakly coupled gravitational theories in $(D + 1)$-dimensional AdS space-time with strongly coupled conformal field theories (CFTs) defined on the $D$-dimensional boundary of AdS. In particular, it gives support to study the transport coefficients from hydrodynamics to the quark-gluon plasma formed at relativistic heavy-ion collisions \cite{Jiang:2017imk}-\cite{Hartnoll:2009sz}.  Shear viscosity $\eta$ \cite{D.T. Son}-\cite{Kovtun:2004de} as one of these coefficients is the most well-known coefficient calculated by this duality, especially fluid-gravity duality\cite{Bhattacharyya:2007vjd}-\cite{Bhattacharya:2011tra}.\\
The Kovtun-Son-Starinets (KSS)  bound states that the ratio $\eta/s$ has a lower bound,
$\frac{\eta }{s} \ge \frac{1 }{4\, \pi }$, for all relativistic quantum field theories \cite{Son:2007vk},\cite{Kovtun:2004de} and can be interpreted as the Heisenberg uncertainty principle \cite{Policastro:2001yc},\cite{Kovtun:2012rj}  where $\eta$ and s are the shear viscosity and entropy density, respectively. However, this conjecture violates for some theories like the Gauss-Bonnet  gravity \cite{Brigante:2008gz}, the Gauss-Bonnet gravity in presence of $U(1)$ gauge field \cite{Ge:2008ni},\cite{Ge:2009eh}, Horndeski theory \cite{Feng:2015oea}, massive gravity for $c_i<0$ \cite{Sadeghi:2020lfe}, scalar-tensor gravity\cite{Bravo-Gaete:2020lzs}-\cite{Bravo-Gaete:2022lno} and anisotropic black brane \cite{Mamo:2012sy}.\\
Observational cosmology shows the accelerated expansion of the Universe \cite{SupernovaSearchTeam:1998fmf},\cite{SupernovaCosmologyProject:1998vns}. Therefore, the Universe at present is becoming filled with a strange substance. This outcome is confirmed by the measurement of the Cosmic Microwave Background (CMB) using PLANCK space Satellite \cite{Planck:2013pxb}. It is believed that this expansion is due to a substantial negative pressure arising from the dark energy that is up to 70$\%$ of the total energy of the Universe. There are two sources of this negative pressure: one is cosmological constant and the other is the so called quintessence \cite{Wang:1999fa},\cite{Tsujikawa:2013fta} which acts as a repulsive force against the gravity. Quintessence is described by a scalar field and a parameter $\omega$, defined as the ratio of the pressure to the energy density of the dark energy. Black hole solution surrounded by the quintessence introduced in \cite{Kiselev:2002dx}.\\
String theory in which the fundamental ingredient of our universe is assumed to be one-dimensional strings, instead of particles, tries to merge the gravitational theory and quantum mechanics. An extension of this idea is to consider a cloud of strings \cite{Letelier:1979ej}-\cite{Ranjbari:2019ktp} and study its possible
measurable effects on long range gravitational fields of some
sources, shuch as black holes. Letelier \cite{Letelier:1979ej} has tried to extend the idea of a string to a black hole, possibly to reveal some properties of the blac khole. Ghosh and et al. obtained a generalization for third-order Lovelock gravity \cite{Ghosh:2014pga}, and Herscovich and et al. \cite{Herscovich:2010vr} obtained the solution for Einstein-Gauss-Bonnet theory in the Letelier spacetime.\\ 
General relativity (GR) provides the standard description of gravity and Gauss-Bonnet gravity is an extension gravity theory. Recently a new proposal has been offered wherein GB term is made  in $4D$ dimentions\cite{Glavan:2019inb}. In that ,the GB coupling is scaled as $\alpha \to \frac{\alpha}{D-4}$ and thereby canceling out $(D-4)$ factor in the equation, and then taking the limit $(D\to4)$. This results into an effective equation in $4D$ which is in fact the Einstein-Gauss-Bonnet (EGB)  equation written in $D=4$. Then it could be solved in spacetime with some specific symmetries for different situations, black holes and cosmology.\\
In this paper, we consider $4D$ Einstein-Gauss-Bonnet-Yang-Mills theory (EGBYM) in the presence of string cloud and quintessence. Then, by calculation of $\frac{\eta}{s}$ via fluid-gravity duality we study the effects of quintessence, string cloud and Yang-Mills charge on the field theory dual side.\\
%--------------------------------------------------------------------------
\section{$4D$ AdS Einstein-Gauss-Bonnet-Yang-Mills in presence of string cloud and quintessence black brane}
\label{sec2}

\indent The action of AdS Einstein-Gauss-Bonnet-Yang-Mills in the presence of string cloud and quintessence is:
\begin{equation}\label{Action}
I =\frac{1}{16\pi }\int{d^Dx\sqrt{-g}\Bigg[R-2 \Lambda+\frac{\alpha'}{D-4}\mathcal{G}+F^{(a)}_{\mu \alpha }F^{(a)\mu \alpha}\Bigg]}+S_{\text{cs}}+S_{\text{quint}},
\end{equation}
 where $ R $ is the scalar curvature, $\Lambda=\frac{-(D-1)(D-2)}{2 l^2}$ , $\alpha'$ is a (positive) Gauss-Bonnet coupling constant with dimension $(\text{length})^2$ , $  \mathcal{G}=R^2-4R_{\mu \nu}R^{\mu \nu}+R_{\mu \nu \rho \sigma }R^{\mu \nu \rho \sigma} $ , $F^{(a)}_{\mu \nu } =\partial _{\mu } A^{(a)}_{\nu } -\partial _{\nu } A^{(a)}_{\mu } -i[A^{(a)}_{\mu }, A^{(a)}_{\nu }]$ is the Cartan subalgebra of $SU(2) $ Yang-Mills  field strength tensor in which the gauge coupling constant is 1, $A_{\nu }$'s are the Cartan subalgebra of the $SU(2)$ gauge group Yang-Mills potentials.\\
%$S_{\text{quint}}$ ,the last term in (\ref{Action}), is the action for quintessence matter. A general action for quintessence in D-dimensional space-time is,
%\begin{equation}\label{Quin}
%S_{\text{quint}}= \frac{1}{2\kappa}\int{d^Dx\sqrt{-g}\Bigg[-\frac{1}{2}(\vec{\nabla} \phi)^2-V(\phi)\Bigg]},
%\end{equation}
%Because of simplicity and preserving space-time symmetry, quintessence is described by an interacting scalar field. 
By considering the following ansatz for the metric, 
\begin{equation}\label{metric1}
ds^{2} =-f(r)N^2(r)dt^{2} +\frac{dr^{2}}{f(r)} +r^2d\Omega^2_{2,\kappa},
\end{equation}where $\kappa = -1,0,1 $ and where  
$\Omega_{D-2}$ is the volume of unit $(D-2)$-sphere.\\
$f(r)$ is defined by a new variable $\phi(r)$,
\begin{equation}\label{f}
f(r)=\kappa-r^2\phi(r).
\end{equation}
The action of cloud of string is as follows,\\ 
\begin{equation}\label{CS}
S_{\text{cs}}=\int_{\Sigma} P \sqrt{-\gamma} d\tau d\sigma,
\end{equation}
where $P$ is non-negative constant related to string tension, $(\tau, \sigma)=(\lambda^0,\lambda^1)$ is a coordiantes of worldsheet of string with $\lambda^0$ and $\lambda^1$ are timelike and spacelike parameters
respectively, $\gamma_{ab}$ is the induced metric on the world sheet is as follows,
 \begin{equation}
\gamma_{ab}=g_{\mu \nu}(x) \frac{\partial x^{\mu}}{\partial \lambda^a}\frac{\partial x^{\nu}}{\partial \lambda^b},
 \end{equation}
and $\gamma=\text{det} \gamma_{ab}\nonumber$ is the determinant of the induced metric.\\ 
The bivector of string worldsheet $\Sigma$ is given by,\\
\begin{equation}
{\Sigma}^{\mu \nu}={\epsilon}^{a b} \frac{\partial x^{\mu}}{\partial \lambda^a}\frac{\partial x^{\nu}}{\partial \lambda^b},
\end{equation}
where ${\epsilon}^{a b}$ is the Levi-Civita tensor in two dimensions, $\epsilon^{01} = -\epsilon^{10} = 1$.\\
Therefore, the action of cloud of string (\ref{CS}) can be written as,
\begin{equation}
S_{\text{cs}}=P \int_{\Sigma}\sqrt{-\frac{1}{2}{\Sigma}^{\mu \nu} {\Sigma}_{\mu \nu}} d\tau d\sigma,
\end{equation}
The non-zero components of the bivector $\Sigma$ is $\Sigma^{tr}=-\Sigma^{rt}=-\frac{a}{P r^{D-2}}$. Where $a$ is an integration constant which is related to the cloud of string \cite{Letelier:1979ej}-\cite{Ranjbari:2019ktp}.\\
The action of quintessence is introduce by Kiselev \cite{Kiselev:2002dx} and written in terms of scalar field as follows,
\begin{equation}
S_{\text{quint}} =\frac{1}{16\pi G }\int{d^Dx\sqrt{-g}\Big[-\frac{1}{2}(\vec{\nabla} \phi)^2-V(\phi)\Big]}.
\end{equation}
The energy-momentum tensor of the quintessence dark energy in $D$ dimensions can be described by
\cite{Chen:2008ra} ,
\begin{align}
T_{t}\,^{t}&=T_{r}\,^{r}=\rho=\frac{\omega \alpha (D-1) (D-2) }{4 r^{(D-1)  (\omega +1)}},\\
T_{x_i}\,^{x_i}&=\frac{\rho}{D-2}((D-1) \omega+1),\,\, i=(1,...,D-2),
\end{align} 
where $\omega$ ($-1<\omega<-\frac{D-3}{D-1}$) and $\alpha$ are the quintessential state parameter and the positive normalization constant related to the density of quintessence, respectively.\\
By variation of the action with respect to $A^{(a)}_{\mu }$ we have, \\
 \begin{eqnarray}\label{EOM-YM}
\nabla_{\mu }F^{(a)\mu \nu } =0,
\end{eqnarray}
For solving Eq.(\ref{EOM-YM}), we consider the ansatz for the gauge field as follows\cite{Shepherd:2015dse},
\begin{equation}\label{background}
{\bf{A}}^{(a)} =\frac{i}{2}h(r)dt\begin{pmatrix}1 & 0 \\ 0 & -1\end{pmatrix},
\end{equation}
so,
\begin{equation}
h(r)=  C_2+Q\int^{r}\frac{ 1}{u^{D-2}}du,
\end{equation}
\begin{equation}
h'(r)= \frac{ Q}{r^{D-2}},
\end{equation}
By plugging the  ansatz \ref{metric1} into the action  Eq.(\ref{Action}) yields,
\begin{align}\label{EOMI}
 &I =\frac{\Omega_{D-2}(D-2)}{16\pi }\nonumber\\&\int{dt dr N(r)\Bigg[r^{D-1}\phi\bigg(1+\alpha'(D-3)\phi\bigg)+\frac{r^{D-1}}{l^2}+\frac{2Q^2 r^{3-D}}{(D-3)(D-2)}-\frac{2 a r}{(D-2)}-\frac{\alpha}{r^{\omega (D-1)}} \Bigg]'},
\end{align}
in which $'$ denotes derivative with respect to $r$. Equation of motion is given by variation
of $\delta N(r)$ \cite{Myers:2010ru}. The function $ \phi $ is given by solving for the roots of a quadratic polynomial,
\begin{equation}\label{phi}
\phi+\alpha'(D-3)\phi^2=\frac{16 \pi M}{(D-2)r^{D-1}\Omega_{D-2}}-\frac{1}{l^2}-\frac{2Q^2 r^{4-2D}}{(D-3)(D-2)}-\frac{2a}{(D-2)r^{D-2}}+\frac{\alpha}{r^{(\omega+1) (D-1)}},
\end{equation}
where $M$ is a constant.\\
By solving Eq.(\ref{phi}) and plugging $\phi$ into Eq.(\ref{f}) we will have,
\begin{equation}
\begin{split}
& f(r)=\kappa-\frac{r^2}{2\alpha'(D-3)}\Bigg[-1\pm \\ &\sqrt{1-4(D-3)\alpha'\bigg(\frac{1}{l^2}+\frac{2Q^2 r^{4-2D}}{(D-3)(D-2)}-\frac{16 \pi M r^{1-D}}{(D-2)\Omega_{D-2}}+\frac{2 a r^{2-D} }{D-2}+\frac{\alpha}{r^{(D-1)(\omega+1)}}\bigg)}\Bigg].
\end{split}
\end{equation}
 The line elements of planar black branes can be written as,\\
\begin{equation}\label{metric}
ds^{2} =-H(r)N(r)^2dt^{2} +\frac{ dr^2 }{H(r)} +\frac{r^2}{l^2} \sum_{i=1}^{D-2}dx_i^2.
\end{equation}
That's easy to show $ N(r) $ is constant by variation of $ \phi(r) $ from Eq.(\ref{EOMI}).
\begin{equation}\label{H}
\begin{split}
	& H(r)=\frac{r^2}{2\alpha'(D-3)}\Bigg[1- \\ &\sqrt{1-4(D-3)\alpha'\bigg(\frac{1}{l^2}+\frac{2Q^2 r^{4-2D}}{(D-3)(D-2)}-\frac{16 \pi M r^{1-D}}{(D-2)\Omega_{D-2}}+\frac{2 a r^{2-D} }{D-2}+\frac{\alpha}{r^{(D-1)(\omega+1)}}\bigg)}\Bigg],
\end{split}
\end{equation}
where $M$ and $Q$ are integration constants proportional to the mass and charge of the black hole
respectively given by the following formulas:\\
 \begin{equation}
 \begin{split}
&M=\frac{(D-2)\Omega_{D-2}}{16 \pi}m,\\
&Q^2=\frac{(D-2)(D-3)}{2}q^2.\\
\end{split}
 \end{equation}
By defining new parameters as below,\\
 \begin{equation}
\begin{split}
&\frac{(D-3)\alpha'}{l^2}=\lambda_{gb},\\
&\frac{2 a  }{D-2}=A,
\end{split}
\end{equation} 
 
and substituting them into the Eq.(\ref{H}) we will have,
\begin{equation}\label{H1}
H(r)=\frac{r^2}{2\lambda_{gb} l^2}\Bigg[1- \sqrt{1-4\lambda_{gb} \bigg(1+\frac{q^2l^2}{r^{2D-4}}-\frac{ m l^2}{r^{D-1}}+\frac{A l^2}{r^{D-2}}+\frac{\alpha l^2}{r^{(D-1)(\omega+1)}}\bigg)}\Bigg].
\end{equation}
Black hole has an event horizon, so by applying the condition $H(r_+)=0$ we can find $m$ as follows,
\begin{equation}
m=\frac{r_+^{D-1}}{l^{2}}+\frac{q^2}{r_+^{D-3}}+A r_++\frac{\alpha l^2}{r_+^{(D-1)\omega}}.
\end{equation}
By plugging $m$ in Eq.(\ref{H1})
\begin{equation}
H(r)=\frac{r^2}{2\lambda_{gb} l^2}\Bigg[1-\sqrt{1-4\lambda_{gb}\Bigg(1-(\frac{r_+}{r})^{D-1}+l^2 q^2 \bigg(\frac{1}{r^{2D-4}}-(\frac{r_+}{r})^{D-1}\frac{1}{r_+^{2D-4}}\bigg)+\mathcal{P}\bigg)\Bigg)}\Bigg],
\end{equation}
where,
\begin{equation}
\mathcal{P}=A l^2\bigg(\frac{1 }{r^{D-2}}-(\frac{r_+}{r})^{D-1}\frac{1}{r_+^{D-2}}\bigg)+\frac{\alpha l^2}{r^{(D-1)}}\bigg(\frac{1}{r^{(D-1)\omega}}-\frac{1}{r_+^{(D-1)\omega}}\bigg).
\end{equation}
In AdS/CFT correspondence, the speed of light in the boundary CFT is simply $c=1$. So that, for the black brane solution in the asymptotic region we have $\lim_{r \to 0}{N(r)^2 H(r)}=1$ to recover a causal boundary. By applying this criterion we will have,\\
\begin{equation}
N^2=\frac{1+\sqrt{1-4\lambda_{gb}}}{2}.
\end{equation}
The temperature and the Hawking-Bekenstein entropy density follow as,
\begin{eqnarray}
&T=\frac{1}{4\pi \sqrt{g_{rr} g_{tt} } } \partial_r g_{tt}|_{r=r_+ } =\frac{Nr_{+}}{4\pi l^2}\Big[(D-1)+\frac{Al^2}{r_+^{D-2}}+\frac{(D-3)l^2q^2}{r_+^{2D-4}}-\frac{(D-1)\alpha \omega}{r_+^{(D-1)(\omega+1)}}\Big],\\
&s=\frac{4\pi}{V}\int d^{3}x \sqrt{-g}=4\pi\left(\frac{r_+}{l}\right)^{3}.
\end{eqnarray}

%--------------------------------------------------------------------------

\section{$\frac{\eta}{s}$ of this solution}
 \label{sec4}
By introducing new variables $ z=\frac{r}{r_+} $ , $ \omega=\frac{l^2}{r_+}\tilde{\omega} $ , $ k_3=\frac{l^2}{r_+ ^2}\tilde{k}_3 $, , $ \tilde{H}(z)=\frac{l^2}{r_+ ^2}H $, The line element in terms of new variables is as follows,
\begin{equation}\label{metricGB3}
ds^{2} =-N^2\tilde{H} dt^{2} +\frac{ dz^2 }{r_+^2\tilde{H}} +\frac{r_+^2 z^2}{l^2} \sum_{i=1}^{D-2}dx_i^2,
\end{equation}
where
\begin{equation}
\tilde{H}(z)=\frac{z^2}{2\lambda_{gb} }\Bigg[1-\sqrt{1-4\lambda_{gb}\Bigg(1-\frac{1}{z^{D-1}}+\frac{l^2 q^2}{r_+^{2D-4}} \bigg(\frac{1}{z^{2D-4}}-\frac{1}{z^{D-1}}\bigg)+\mathcal{P}\Bigg)}\Bigg],
\end{equation}
\begin{equation}
\mathcal{P}=\frac{A l^2}{r_+^{D-2}}\bigg(\frac{1 }{z^{D-2}}-\frac{1 }{z^{D-1}}\bigg)+\frac{\alpha l^2}{r_+^{(D-1)(\omega+1)}}\bigg(\frac{1}{z^{(D-1)\omega}}-\frac{1}{z^{(D-1)(\omega+1)}}\bigg).
\end{equation}
To calculate the shear viscosity, we perturb the background metric by $ h^{y}_{x} $. We call the perturbed part of metric  $\psi=h^{y}_{x}$ \cite{Ref26,Ref27}. 
\begin{equation}\label{metricGBYMCS}
ds^{2} =-N^2\tilde{H} dt^{2} +\frac{ dz^2 }{u^2\tilde{H}} +\frac{ z^2}{u^2 l^2}\Bigg( \sum_{i=1}^{D-2}dx_i^2+2\psi(t,\vec{x},z)dx_1dx_2\Bigg),
\end{equation}
where $u=\frac{1}{r_+^2}$.\\
Using Fourier decomposition, 
\begin{equation}
\psi(t,x_i,z)=\int{\frac{d^{D-1}k}{(2\pi)^{D-1}}e^{-i\tilde{\omega}t+\tilde{k_i} x_i}}\psi(k,z).
\end{equation}
We substitute the perturbed metric Eq.(\ref{metricGBYMCS}) in Eq.(\ref{Action}) and expand the action up to second order of $\psi$. Finally, the equation of motion for $\psi(z)$ can be obtained by variation of  the perturbed action with respect to $\psi$ as follows,\\
\begin{equation}\label{Mod}
(K \psi')'+\omega^2K_1\psi-k_i^2K_2\psi=0,
\end{equation}
where
 \begin{align}\label{54}
  &K=\frac{1}{16 \pi}\sqrt{-g}g^{zz}g^{x_i x_i}K_3(z),\nonumber\\
  &K_1=-\frac{1}{16 \pi}\sqrt{-g}g^{tt}K_3(z), \nonumber\\
  &K_2=\sqrt{-g}g^{x_i x_i}K_4(z), \nonumber\\
  &K_3(z)=1-\frac{2\lambda_{gb}}{D-3}[z^{-1} \tilde{H}'+z^{-2} (D-5) \tilde{H}], \nonumber\\
   &K_4(z)=1-\frac{2\lambda_{gb}}{(D-3)(D-4)}[\tilde{H}''+ (D-5)(D-6)z^{-2} \tilde{H}+2(D-5)z^{-1}\tilde{H}'], \nonumber\\
   \end{align} 
in which $'$ denotes derivative with respect to $z$. The factors $(D-5)$ and
$(D-6)$ in the expression of $K_4(z)$ is for $D>5$ of the Gauss-Bonnet theory.\\
The Green function is,
\begin{equation} 
G_{x_i x_j, x_i x_j}=\frac{K\psi'}{\psi}\, .
\end{equation}
The shear viscosity is calculated by Green-Kubo formula as the following,
\begin{equation}
\eta_{x_i x_j, x_i x_j}=\frac{-G_{x_i x_j, x_i x_j}}{i \omega},
\end{equation}
Now we write Eq.(\ref{Mod}) in terms of the shear viscosity, 
\begin{equation}
\partial_{z}\eta_{x_i x_j, x_i x_j}=(\frac{\eta^2_{x_i x_j, x_i x_j}}{K}-K_1)+\frac{i}{\omega}K_2k_i^2,
\end{equation}
and then we can compute the shear viscosity by requiring horizon regularity as follows,
\begin{equation}
\eta_{x_i x_j, x_i x_j}=\sqrt{K K_1}\Bigg|_{z=1}=\frac{1}{16\pi}\frac{r_+^{D-2}}{l^{D-2}}  \left(1-\frac{2 \lambda  }{D-3}\tilde{H}'(1)\right),
\end{equation}
The ratio of the shear viscosity to the entropy density for 4D charged black hole solutions
in Gauss-Bonnet gravity in the presence of string cloud and quintessence is then\\
\begin{equation}
\frac{\eta}{s}=\frac{1}{4\pi }[1-\frac{2 \lambda  }{D-3}\tilde{H}'(1)],
\end{equation}
\begin{equation}
\frac{\eta}{s}=\frac{1}{4\pi }\Bigg[1-\frac{2 \lambda  }{D-3}\Bigg((D-1)-(D-3) l^2 q^2 r_+^{4-2 D}+A l^2 r_+^{2-D}+\alpha l^2 (D-1) r_+^{(1-D)(1+\omega)}\Bigg)\Bigg].
\end{equation}
For $\lambda \to 0$ this value is $\frac{\eta}{s}=\frac{1}{4\pi }$. It means that the KSS bound is saturated for Einstein-Hilbert gravity.\\
For $q=A=\alpha=0$ we will have,
\begin{equation}
\frac{\eta}{s}=\frac{1}{4\pi }\big[1-2\frac{(D-1)}{(D-3)}\lambda \big],
\end{equation}
which is consistent with the literature\cite{Brigante:2007nu}.\\
In the limit $D \to 4$, we obtain
\begin{equation}
\frac{\eta}{s}=\frac{1}{4\pi }\Bigg[1-2 \lambda \Bigg(3- l^2 q^2 r_+^{-4}+A l^2 r_+^{-2}+3\alpha l^2  r_+^{-3(1+\omega)}\Bigg)\Bigg].
\end{equation}
For $A=\alpha=0$ the result is the same as \cite{Ge:2020tid},
\begin{equation}
\frac{\eta}{s}=\frac{1}{4\pi }\Bigg[1-2 \lambda \Big(3- l^2 q^2 r_+^{-4}\Big)\Bigg].
\end{equation}
%--------------------------------------------------------------------------
\section{Conclusion}

\noindent In summary, we have investigated the effect of string cloud , quintessence and Yang-Mills charge in $4D$ Gauss-Bonnet gravity on the field theory dual side by the fluid-gravity duality. Our result shows that theses three matters are affected in dual of Gauss-Bonnet gravity but they have no contribution in dual of Einstein-Hilbert gravity.   Therefore,  the KSS conjecture \cite{Kovtun:2004de} is saturated in the limit $\lambda \to 0$ . This conjecture tells us that the ratio $\eta/s$ has a lower bound, $\frac{\eta }{s} \ge \frac{\hbar }{4\, \pi \, k_{B} } $, for all relativistic quantum field theories at finite temperature without chemical potential \cite{Policastro:2001yc} and can be interpreted as the Heisenberg uncertainty principle \cite{Son:2007vk}. String cloud \cite{Sadeghi:2019muh} and quintessence \cite{Sadeghi:2020lfe} in Einstein massive gravity does not contribute to the $\frac{\eta}{s}$. However, this conjecture violates for higher derivative theories of gravity like the AdS Einstein-Gauss-Bonnet-Yang-Mills in the presence of string cloud and quintessence as we confirmed in section 3 . Our outcome shows that the coupling of this model is scaled as $\lambda \to \lambda \Big((D-1)-(D-3) l^2 q^2 r_+^{4-2 D}+A l^2 r_+^{2-D}+\alpha l^2 (D-1) r_+^{(1-D)(1+\omega)}\Big)$.
%--------------------------------------------------------------------------

\vspace{1cm}
\noindent {\large {\bf Acknowledgment} }  Author would like to thank Xian-Hui Ge , M.H. Abbasi ,and the referees of MPLA for useful comments and suggestions.

%--------------------------------------------------------------------------

\vspace{1cm}
\noindent {\large {\bf Data Availability } } This manuscript has no associated data.

%--------------------------------------------------------------------------

\end{document}